\documentclass[twocolumn,amsmath,floats,showpacs,nofootinbib]{revtex4}
\usepackage[usenames]{color}
\usepackage{graphicx}
\usepackage{epstopdf}
\usepackage{textcomp}
\usepackage{hyperref}
\usepackage{multirow}
\usepackage{tikz}
\usepackage{rotating}
\hypersetup{colorlinks=true}

\begin{document}

\title{Tunneling measurements of the electron-phonon interaction in $\rm Ba_{1-x}K_xBiO_3$}

\author{P. Samuely}
\affiliation{Max-Planck-Institut f$\ddot{u}$r Festk$\ddot{o}$rperforschung, Hochfeld-Magnetlabor, F-38042 Grenoble Cedex, France
and Institute of Experimental Physics, Slovak Academy of Sciences, CS-04353 Ko$\check{s}$ice, Slovakia}
\author{N.L. Bobrov}
\affiliation{Max-Planck-Institut f$\ddot{u}$r Festk$\ddot{o}$rperforschung, Hochfeld-Magnetlabor, F-38042 Grenoble Cedex, France;
Institute of Experimental Physics, Slovak Academy of Sciences, CS-04353 Ko$\check{s}$ice, Slovakia
and Institute for Low Temperature Physics and Engineering, Ukrainian Academy of Sciences, 310164 Kharkov, Ukraine
Email address: bobrov@ilt.kharkov.ua}
\author{A.G.M. Jansen, P.Wyder}
\affiliation{Max-Planck-Institut f$\ddot{u}$r Festk$\ddot{o}$rperforschung, Hochfeld-Magnetlabor, F-38042 Grenoble Cedex, France}
\author{S.N. Barilo, and S.V. Shiryaev}
\affiliation{Institute of Physics of Solids and Semiconductors, Belarus Academy of Sciences, 220726 Minsk, Belarus}

\published {\href{https://doi.org/10.1103/PhysRevB.48.13904}{Phys. Rev. B}, \textbf{48}, 13904 (1993)}
\date{\today}

\begin{abstract}The conductance curves of point-contact tunnel junctions between Ag and $\rm Ba_{1-x}K_xBiO_3$ ($x\simeq 0.4$) reveal a BCS behavior with low leakage current at zero voltage and some broadening of the superconducting-gap structure. In the energy range above the superconducting energy gap, the structure in the voltage dependence of the second derivative $d^2V/dI^2$ of the voltage with respect to the current of the tunnel junction has been investigated in detail in magnetic fields up to $10\ T$. While part of this structure is rapidly changing in a magnetic field, three reproducible peaks in $d^2V/dI^2(V)$ remain stable up to the transition temperature from the superconducting to the normal state with only additional broadening in the applied magnetic field. An analysis of this structure in terms of strong-coupling effects yields the spectral function $\alpha^2F$ for the electron-phonon interaction. The obtained spectral weight in the energy region 20-70~$meV$ points to the importance of the oxygen optical modes in the electron-phonon coupling for the superconductivity of $\rm Ba_{1-x}K_xBiO_3$.

\end{abstract}

\maketitle
\section{INTRODUCTION}
Among the fundamental questions in high-$T_c$ superconductors, a description of the normal state of the superconducting oxides in the framework of a Fermi liquid remains a central issue \cite{1}. Within the Fermi-liquid description, the BCS theory of superconductivity predicts an energy gap in the density of states of quasi-particle excitations. In an experimental approach of this question, tunneling spectroscopy has been a very decisive probe for the classical superconductors via a direct measurement of the quasiparticle density of states $N$ of the superconductor (proportional to the conductance $dI/dV$ of the tunnel junction). The observed sharp BCS singularity in the quasiparticle excitation spectrum clearly demonstrates the existence of the superconducting energy gap and, moreover, strongly supports the existence of a Fermi surface. Tunneling experiments have also clarified the electron-pairing mechanism in the classical superconductors, namely the electron-phonon interaction.

Despite the effort of many groups, the tunneling data on the high $T_c$'s remain ambiguous and controversial \cite{2}. Discrepancies in the tunneling behavior of high-$T_c$ superconductors with respect to the BCS theory are often believed to reflect intrinsic properties of the new materials as, for example, the quasi two-dimensional character of the lattice with the stacked $ab$ planes and the spin fluctuations leading to magnetic moments. The former can account for the inhomogeneity of the order parameter which can cause a (broadened) multipeak structure in the tunneling conductance \cite{3}, while the latter can explain the linear background \cite{4}. In the bismuthates both features are absent, i.e., the superconducting $\rm Ba_{1-x}K_xBiO_3$ has a cubic structure and is diamagnetic in the normal state. On the other hand the system is a perovskite superconducting oxide. Hence, this simplified system can throw some light on the more complex copper oxides.

In the first tunneling measurements on $\rm Ba_{1-x}K_xBiO_3$ (Refs.\cite{5} and \cite{6}) tunneling conductances with a finite zero-bias signal and smeared gap-like anomalies (typical phenomena for the copper oxides) were observed despite of the simplicity of the system with a cubic and nonmagnetic structure. In these experiments an indication of the phonon mechanism of superconductivity was found in the observed nonlinearities of the tunneling conductance in the voltage (energy) range of the phonons. Conclusive BCS behavior was found in later work by the same group \cite{7} and by others \cite{8,9} in tunneling data with a complete superconducting-energy-gap structure. In these experiments \cite{7,8} the electron-phonon coupling in the $\rm Ba_{1-x}K_xBiO_3$ tunneling curves was also investigated.

McMillan and Rowell \cite{10} have introduced a method to recover in a self-consistent way the spectral function $\alpha^2F(\omega)$ for the electron-phonon interaction from the measured density of states using the integral equations describing superconductivity with a strong electron- phonon coupling \cite{11}. Here $\alpha^2F(\omega)$ is the convoluted product of the phonon density of states $F(\omega)$ and the matrix element
squared for the electron-phonon coupling. Twice the integral of the function $\alpha^2F(\omega)/ \omega$ over the phonon frequencies $\omega$ defines the electron-phonon coupling constant $\lambda$, which, together with the Coulomb repulsion term $\mu^*$, determines the superconducting transition temperature $T_c$.

The electron-phonon-interaction mechanism for the superconductivity in $\rm Ba_{1-x}K_xBiO_3$ has been studied theoretically by Shirai, Suzuki, and Motizuki \cite{12} using the realistic electronic bands of $\rm BaBiO_3$. The calculated electron-phonon interaction function $\alpha^2F(\omega)$ has prominent features in the frequency range of the oxygen stretching and/or breathing modes, which are expected to dominantly contribute to the superconductivity \cite{12}. The electron-phonon interaction constant $\lambda$ exceeds 1.0 for compounds with a potassium content less than 0.3 indicating a strong electron-phonon coupling.

In the work of Huang \emph{et al}. \cite{7} evidence has been reported for the predominant electron-phonon coupling in $\rm Ba_{1-x}K_xBiO_3$. Nonlinearities in the tunneling characteristics above the superconducting energy gap have been interpreted as the phonon effect in the superconducting density of states. By means of a modified McMillan-Rowell inversion scheme (accounting for the proximity effect) the Eliashberg function $\alpha^2F(\omega)$ has been obtained with a resemblance to the phonon density of states $F(\omega)$ \cite{13}. The authors also obtained a remarkable agreement between the calculated and experimental $T_c$.

Sharifi \emph{et al}. \cite{8} found the nonlinearities in the tunneling conductance above the energy gap were not reproducible. Moreover, the authors argued that the nonlinearities are not symmetric in the bias voltage but in the current indicating their origin in spurious effects such as microshort-path changes or flux-line motions induced by current.

To elucidate the origin of the nonlinearities in the tunneling conductance, we have measured in detail the voltage dependence of the second derivative of the current-voltage characteristics of tunnel contacts with $\rm Ba_{1-x}K_xBiO_3$ under the influence of external parameters such as temperature and magnetic field.

In our conductance curves of the $\rm Ag-Ba_{1-x}K_xBiO_3$ tunnel junctions we observed the BCS behavior with a pronounced energy gap together with a (sub)linear and slightly asymmetric background with respect to the zero-bias voltage. The superconducting-gap structure disappears at the transition to the normal state. In addition, structure was found in the tunneling conductance at voltages above the energy gap. Part of this structure was stable in an applied magnetic field and was observed for a number of different point contacts. The maxima in the second derivative $d^2V/dI^2$ of the voltage with respect to the current can be related to the peaks in the phonon density of states obtained by neutron-scattering experiments \cite{13}. The McMillan-Rowell inversion scheme has been applied to obtain the Eliashberg function for the electron-phonon interaction. The reasonable agreement with the calculated electron-phonon interaction spectrum and the resemblance with the phonon density of states indicate a significant role of the phonons in the mechanism of superconductivity in $\rm Ba_{1-x}K_xBiO_3$.
\section{EXPERIMENT}
The $\rm Ba_{1-x}K_xBiO_3$ single crystals have a potassium content x ranging from 0.41 to 0.43. x-ray diffraction studies show the monophase structure with the cubic symmetry. The transition to the superconducting state is quite wide with the onset of superconductivity at $T_c^{onset}=28\ K$ and the full superconductivity at $T_c^{end}= 18\ K$, probably associated with the inhomogeneity of the potassium distribution in the crystal.

The point-contact technique was used to make the tunnel junctions. A silver single-crystal needle was pressed on the freshly cleaved, deep-blue colored surface of the crystal at liquid-helium temperature. The inset contained three differential screw mechanisms which allows us to change the position of the tunneling contact as well as the pressure between the tip and the crystal at low temperatures. The measurements were performed in the temperature range from 1.5 to 30~$K$ in magnetic fields up to 10~$T$. The current modulation technique was applied to detect phase-sensitively the first and second derivative of the voltage with respect to the current of the tunnel junction.
\section{RESULTS AND DISCUSSION}
The point-contact junctions use the natural oxides or some degradation on the surface layers of the contacting electrodes to form the tunnel barrier. The barrier is controlled by a smooth adjustment of the pressure between the electrodes. In contrast to the sandwich-type junctions, the characteristics of the interface and the tunnel barrier are not well defined here. On the other hand, the point-contact method is very suitable for studies on the high-$T_c$ superconductors with the known problems of surface degradation and inhomogeneous stoichiometry, because the "best" place can be found by such a local probe on the surface.

\begin{figure}[]
\includegraphics[width=8.5cm,angle=0]{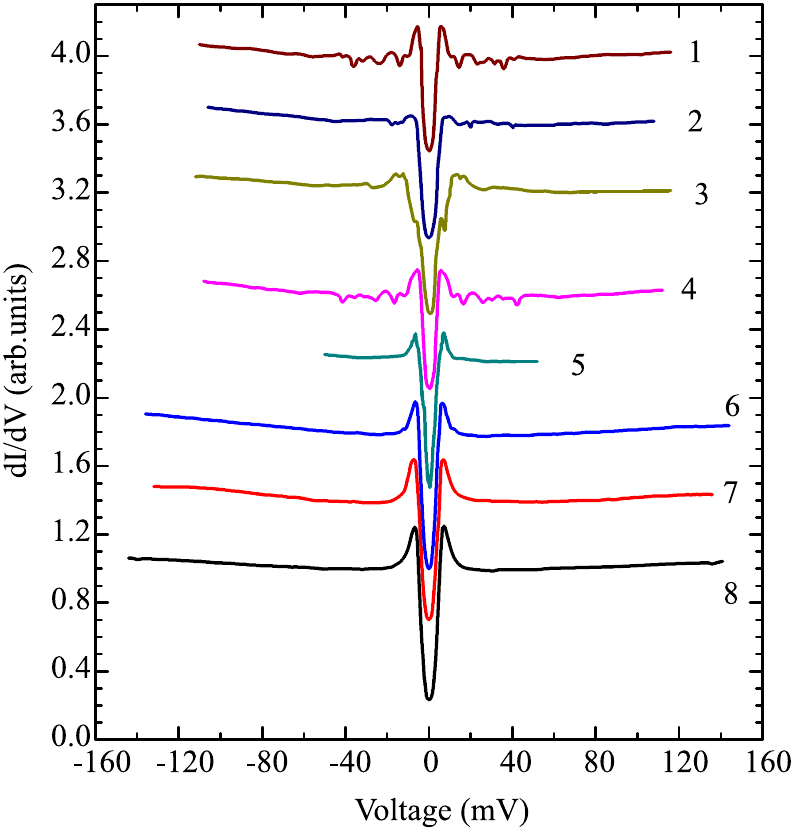}
\caption[]{Tunneling conductance traces $dI/dV$ for different point-contact tunnel junctions $\rm Ag-Ba_{1-x}K_xBiO_3$. All the curves were recorded at T=4.2~$K$. For clarity, the curves are shifted (except the bottom one) in the conductance scale taking the high voltage value as a reference point.}
\label{Fig1}
\end{figure}

A large number of point contacts have been examined in magnetic fields up to 10~$T$ and at various temperatures. The best resolved tunneling spectra were achieved after the first soft touch of the electrodes at 4.2~$K$. Figure \ref{Fig1} displays a variety of observed tunneling conductances $dI/dV$ for different $\rm Ag-Ba_{1-x}K_xBiO_3$ point contacts measured at 4.2~$K$. On each curve a reduced conductance is observed around zero-bias voltage and a maximum in both voltage polarities. The maxima are symmetric with respect to the zero-bias voltage. The relative changes of the conductance around zero bias vary from contact to contact. In the best examples (lower four curves) the conductance at zero bias is about 20\% of the high voltage value. In these cases as will be shown below, the observed zero bias conductance is completely compatible with the width $\Gamma$ of the main peak. This fact points to a low leakage current in these junctions. Most of the current is tunneling through the barrier and the maximum can be related to the superconducting energy gap. In most cases we observed a slightly asymmetric (sub)linear background. The positive bias voltage in the displayed characteristics corresponds to a positively biased Ag electrode.

The upper four curves in Fig.\ref{Fig1} illustrate the kind of spurious effects which could be observed. In these curves, the incompatibility of the zero-bias conductance with the width of the gap-like peaks indicates the presence of leakage currents. Such a phenomenon is normally observed in the point-contact tunnel junctions with high-$T_c$ superconductors. In the voltage range above the main peaks up to 50~$mV$, we often observed dips in the conductance symmetrically positioned with respect to the zero-bias voltage. The dips are very sensitive to the magnetic field. Unlike the main peak in the conductance curves, the dips disappear when a small magnetic field of about 200~$mT$ is applied. The voltage positions of the dips are not reproducible, but change from contact to contact. The dips could originate from a redistribution of the current paths when the critical current is reached in a weak link or crack nearby the contact. Due to the weakness of the superconductivity in these places, the redistribution can also be influenced by a small magnetic field. Two pairs of peaks can be seen in curve 3 with the first one positioned at the usual voltage of the gaplike peak and the other one at twice the voltage. As a possible explanation two parallel contacts can be considered, where beside the expected N-S point contact (N denotes the normal Ag electrode and S the superconducting oxide) there is an S-S contact. The S-S contact could be created with a piece of the superconducting $\rm Ba_{1-x}K_xBiO_3$ sticking on the Ag electrode.

We describe all these anomalous effects only to point out that all of them can also be found in the superconducting copper oxides, see, e.g., Ref.\cite{14}. They are sometimes taken as an indication of unusual intrinsic properties. However, taking into consideration that for the bismuthates tunneling data on high-quality single crystals and epitaxial films \cite{7,8,9} have shown the complete BCS character of the quasiparticle excitation spectrum, all these effects can be referred to as "parasitic" here. The same parasitic or extrinsic origin may be responsible for the anomalies in the tunneling data of the cuprate high-$T_c$ superconductors.
\begin{figure}[]
\includegraphics[width=8.5cm,angle=0]{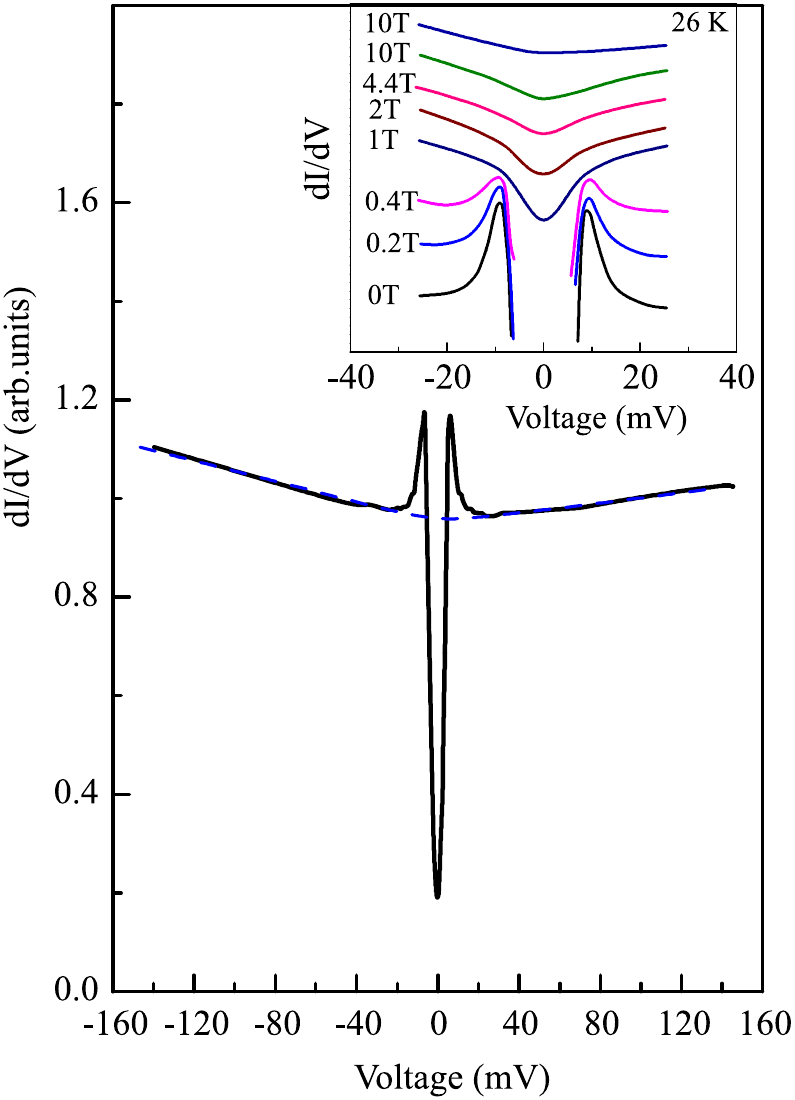}
\caption[]{The tunneling conductance for a $\rm Ag-Ba_{1-x}K_xBiO_3$ tunnel junction (6 in Fig.\ref{Fig1}) in the superconducting state (solid line: $B=0\ T$ and $T= 4.2\ K$) and in the normal state (broken line: $B= 10\ T$ and $T=26\ K$). The inset shows the influence of the magnetic field on the gap-related structure.}
\label{Fig2}
\end{figure}

In Fig.\ref{Fig2} the tunneling conductance of a point contact with low leakage current is shown in the superconducting state (solid line) as well as in the normal state (broken line). The superconducting state conductance was measured at 4.2~$K$. When a magnetic field is applied, there is a pair-breaking effect on the gaplike peak (inset Fig.\ref{Fig2}). The zero-bias structure indicates that a 10-$T$ field is not sufficient to suppress the superconductivity (Nagata \emph{et al}. \cite{15} reported $H_{c2}= 16\ T$ for the upper critical magnetic field). When the temperature was increased up to 26~$K$ the dip disappeared and only the background conductance which we attribute to the normal state remained. This normal-state tunneling curve is displayed in Fig.\ref{Fig2} as the dashed line. As one can see the normal-state conductance fits well the overall background in the superconducting state curve. Small differences should be attributed to possible changes in the point-contact resistance during the magnetic field and temperature sweeps.

\begin{figure}[]
\includegraphics[width=8.5cm,angle=0]{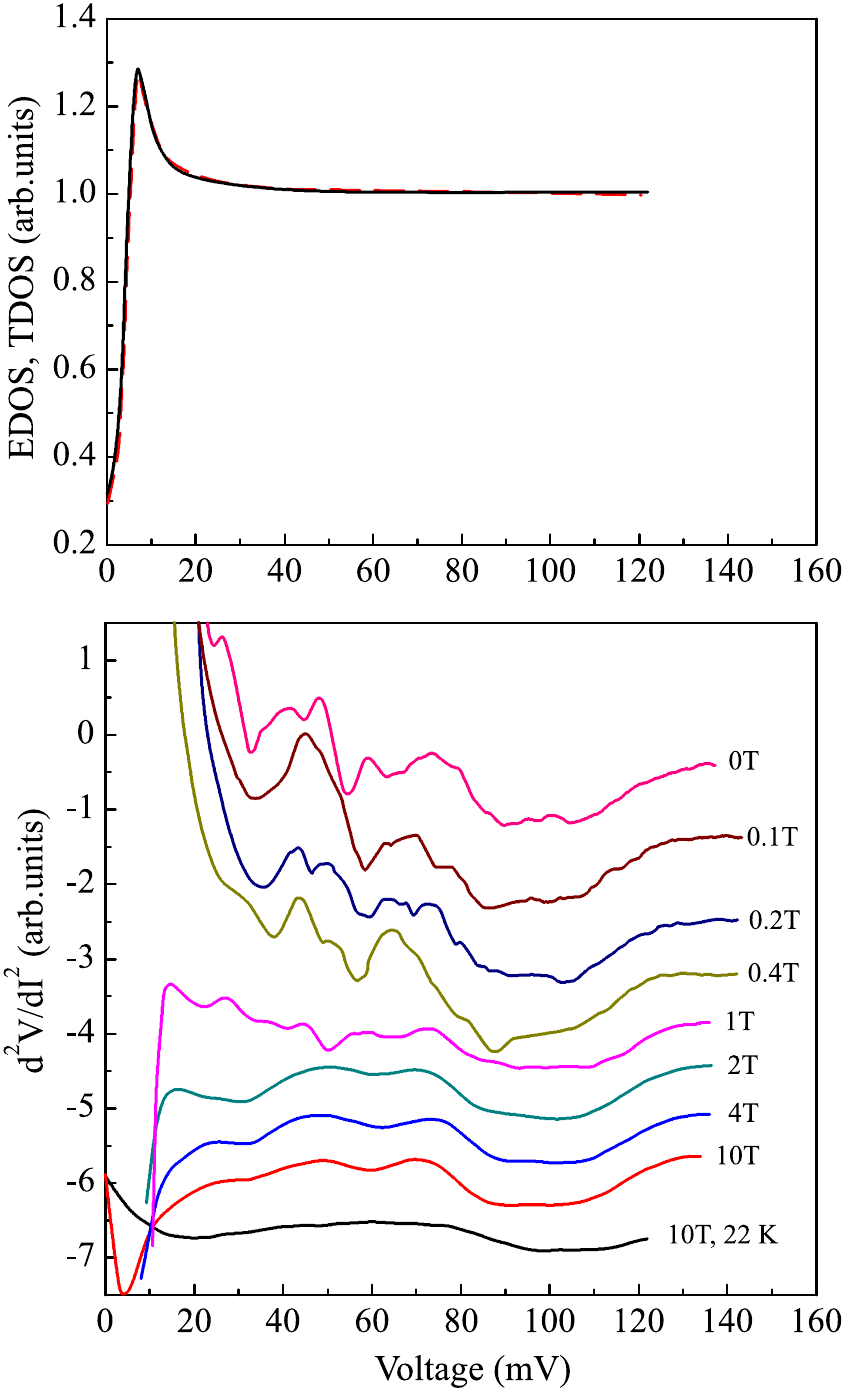}
\caption[]{Upper part: the solid line displays the tunneling conductance of a $\rm Ag-Ba_{1-x}K_xBiO_3$ tunnel junction (7 in Fig.\ref{Fig1}) normalized with respect to the normal state. The dashed line is the calculated life-time broadened density of states with $\Delta= 5\ meV$ and $\Gamma= 1.7\ meV$. Lower part: the second derivative $d^2V/dI^2(V)$ of the voltage with respect to the current of the same tunnel junction measured at different magnetic fields. All the curves were taken at $T=4.2\ K$ except the bottom one ($T=22 K$). Curves are shifted in the vertical direction.}
\label{Fig3}
\end{figure}

In the same way we explored the contact shown in Fig.\ref{Fig3}. The solid line in the upper part represents a normalized tunneling conductance referred to as the experimental density of states (EDOS). The tunneling conductance in the superconducting state at 4.2~$K$ and 0~$T$ was normalized to the tunneling conductance in the normal state at an increased temperature $T = 22\ K$ and at 10~$T$. The dashed line in Fig.\ref{Fig3} represents the theoretically calculated BCS density of states (TDOS) as introduced by Dynes, Narayanamurti, and Garno \cite{16} in a phenomenological model taking into account lifetime broadening of the quasiparticle energies. The calculated density of states TDOS is given by
\begin{equation}
\label{eq__1}
\text{TDOS = Re}\{(V-i\Gamma)/[(V-i\Gamma)^2-\Delta^2]^{1/2}\}
\end{equation}
where $\Delta$ is the superconducting energy gap and $\Gamma$ is the lifetime broadening parameter. A good agreement is obtained between these two curves for the parameters $\Delta= 5\ meV$ and $\Gamma= 1.7\ meV$. Consequently, the $2\Delta/k_BT_c$ ratio is about 4.1 indicating a medium strength of coupling.
By adding a constant to the tunneling conductance in both the superconducting and the normal state in order to take account for any leakage current, we obtain a better agreement between the calculated and experimental curve. However, the very small leakage currents ($10^{-4}$ with respect to the tunneling current at high bias voltage) required demonstrates that leakage current is negligible and that only the superconducting density of states (with some additional broadening) is probed in the tunnel junctions. Generally, the origin of the broadening is unknown and may be a consequence of a surface problem, proximity effect, unusual tunnel barrier, or inhomogeneity. We note that our sample is not perfectly homogeneous. An inhomogeneous order parameter can be expected in the scanned area. It has been shown by Kirtley \cite{2} that a distribution in the order parameter of the probed sample would give the same broadening in the tunneling conductances as the above mentioned lifetime broadening.

Special attention was payed to the tiny structure observed in the tunneling conductance at voltages above the energy gap. The corresponding nonlinearities can be better seen in the directly measured second derivative $d^2V/dI^2$ of the voltage with respect to the current. For sufficient signal-to-noise ratio and resolution, we kept the modulation voltage in the second derivative measurement around 1~$mV$. As mentioned before, in many contacts a structure appears at high voltages in the form of conductance dips due to a spurious critical-current effect \cite{8}. It cannot be excluded that in other junctions this structure appears in a weaker form, thereby undermining any argumentations based on an energy-resolved spectroscopic effect. We remark that similar dips have also been observed by Morales \emph{et al}. \cite{17}. Since the asymmetry of the tunneling conductance of our junction is small we cannot distinguish whether the position of these anomalies is symmetric in the voltage or in the current. The symmetry in the current would prove the spurious origin.

In Fig.\ref{Fig3} we have plotted the second derivative curves at various magnetic fields. For an applied magnetic field above 1~$T$ a very stable structure consisting of three peaks emerges, which remain unchanged up to 10~$T$. The persistency of this structure in a strong magnetic field is not compatible with the before mentioned spurious effects. The peaked structure disappeared by driving the junction to the normal state at a temperature $T = 22\ K$. The thermal broadening of the $d^2V/dI^2(V)$ signal with $5.4\ k_BT$ linewidth \cite{18} is not sufficient to smear all details of the peaks for a temperature $T = 22\ K$. Hence, the observed effect is associated with the superconducting state only.

\begin{figure}[]
\includegraphics[width=8.5cm,angle=0]{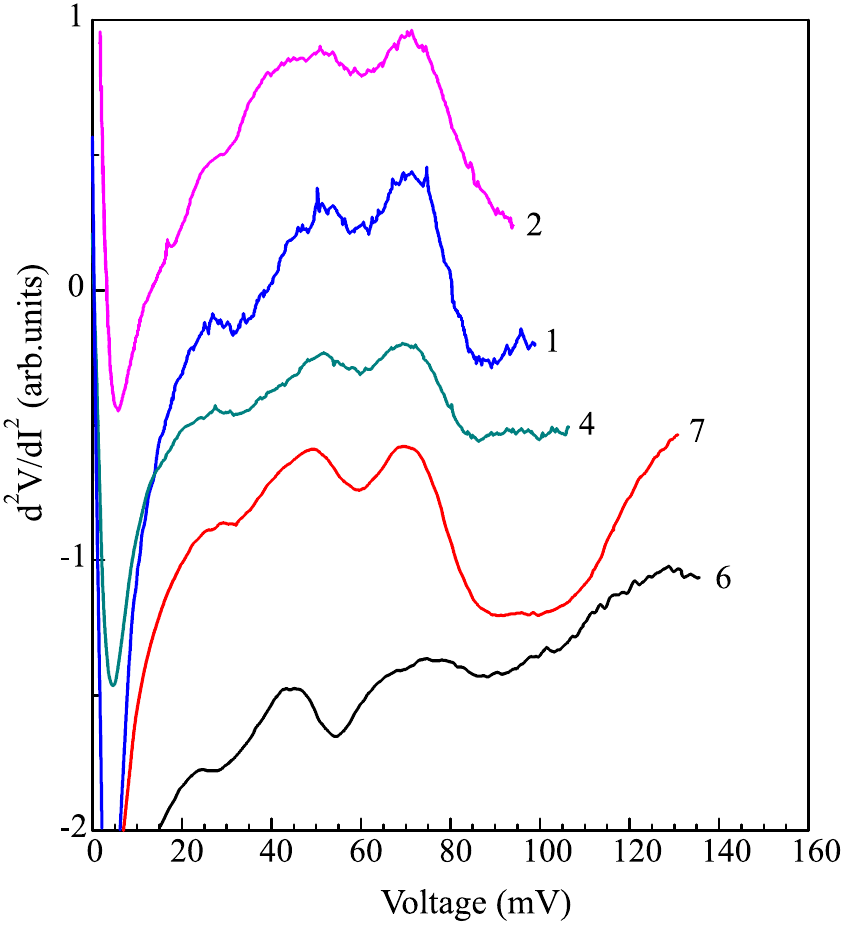}
\caption[]{The measured second derivative $d^2V/dI^2(V)$ of the voltage with respect to the current for different tunnel junctions at $B=10\ T$ and $T=4.2\ K$. The numbering refers to Fig.\ref{Fig1}.}
\label{Fig4}
\end{figure}

In Fig.\ref{Fig4} we have plotted the second derivative at 10~$T$ for different point contacts showing the reproducibility of the three peaks. The tunnel junctions with the identical three peak structure in $d^2V/dI^2$ at higher fields (as shown in Fig.\ref{Fig4}) can show differences in the finer detailed structure at smaller fields, which we relate to the critical current effects. The irreproducible dips in the conductance traces (for examples see Fig.\ref{Fig1}) result in huge S-shaped structures in the second derivative, which hinder the tunnel spectroscopy at small fields. In a magnetic
field this anomalous structure is rapidly changing. Nevertheless, the anomalous structure is suppressed for fields above 1~$T$ while the three-peak structure remains. At this point we stress that we have observed at least two different point contacts (one of them shown in Fig.\ref{Fig3}) with different zero-bias conductances revealing the very same tiny structure above the energy gap at zero magnetic field. These two tunnel junctions have very small spurious effects at zero magnetic field. These two junctions were used in the McMillan-Rowell inversion scheme presented later in the paper.

Relating the three peak structure to the superconducting density of states, the peaks in the $d^2V/dI^2(V)$ characteristics have to be positioned at voltages corresponding to the characteristic phonon (or, more generally, mediating boson) energies $\hbar\omega$ plus the shift with respect to the superconducting energy gap $\Delta$. In our case the three maxima at zero field (Fig.\ref{Fig3}) are positioned around 20, 30-50 and 60-80~$meV$. Accounting for the shift with the superconducting energy gap $\Delta= 5\ meV$, these maxima can be identified with three broad phonon bands centered around 15, 25-45, and 55-65~$meV$ as obtained from the neutron scattering experiment \cite{13}.

A question arises as to why the phonon structure is observable, if the order parameter is itself smeared, e.g., due to stoichiometric inhomogeneity, which would also cause a smearing of the phonon anomalies. For a possible explanation, we refer to the paper of Zasadzinski et \emph{al}. \cite{6}, where pronounced phononlike anomalies in the $\rm Ba_{1-x}K_xBiO_3$ system were observed only for a concentration $x =0.375$, and not for $x =0.5$. This is consistent with theoretical work of Shirai, Suzuki, and Motizuki \cite{12} which describes the dramatic effects on the Eliashberg function $\alpha^2F(\omega)$ with K concentration. Thus the optimal
K concentration, i.e., the optimal order parameter contributes the most.

\begin{figure}[]
\includegraphics[width=8.5cm,angle=0]{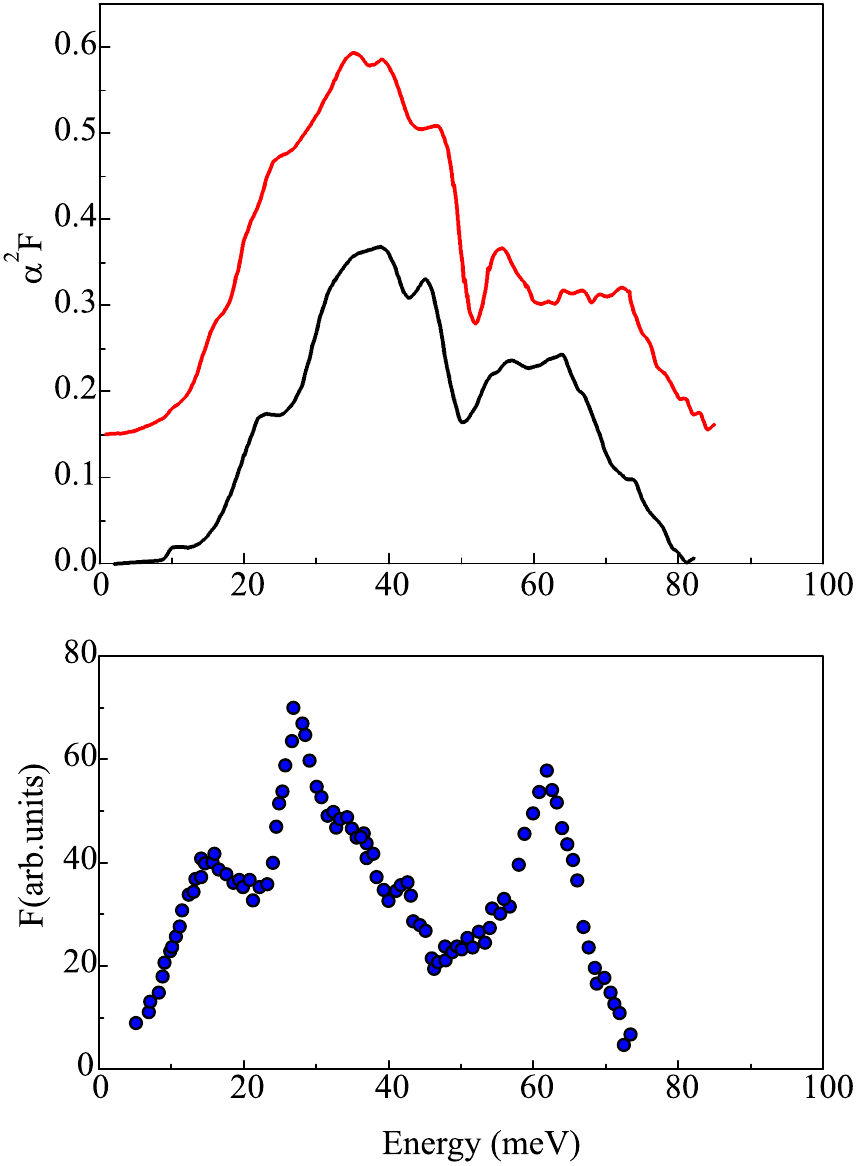}
\caption[]{Upper part: the Eliashberg functions $\alpha^2F(\omega)$ obtained from a McMillan-Rowell analysis of the tunneling data for two different tunnel junctions (7, upper curve, and 8 in Fig.\ref{Fig1}). The upper curve is shifted vertically by 0.15. Lower part: phonon density of states of $\rm Ba_{0.6}K_{0.4}BiO_3$ obtained by inelastic neutron scattering (Ref.\cite{13}).}
\label{Fig5}
\end{figure}

The tunneling data from Fig.\ref{Fig3} and from another point contact yielding the same tiny structure at zero magnetic field have been inverted using the McMillan-Rowell inversion scheme. From the obtained convergence in the iterative procedure, the resulting Eliashberg spectral functions $\alpha^2F(\omega)$ are shown in Fig.\ref{Fig5}. Below 8~$meV$ a quadratic energy dependence of the electron-phonon interaction has been assumed. Besides the Eliashberg function and the electron-phonon interaction parameter $\lambda$, an important result in the McMillan-Rowell analysis is the value of the Coulomb repulsion parameter $\mu^*$ and the calculated superconducting transition temperature $T_c$. We obtained for the two contacts $\lambda=1$, respective 0.7, $\mu^* =0.07$, respective 0 and for the transition temperature $T_c= 30.8\ K$, respective 31.7~$K$.

Concerning the possible observation of the phonon structure in the tunneling data, we rely more on our direct measurements of the second derivative $d^2V/dI^2(V)$ than on the results of the inversion procedure which can be affected by uncertainties in the reduced density of states. The input data of the McMillan-Rowell inversion require the tunneling conductance in the superconducting state normalized with that in the normal state. Although the overall shape of the background in the normal-state tunneling conductance agrees well with the superconducting one (see Fig.\ref{Fig2}), small differences could be found in the high-voltage slope probably related to changes in the point-contact resistance during the magnetic field and temperature sweep. Moreover, especially at low bias voltages, the presence of the broadening parameter $\Gamma$ in the description of the superconducting tunneling conductance yields some
uncertainty in the derived reduced density of states. Both effects influence the distribution of spectral weight in the function $\alpha^2F(\omega)$ resulting from the structures in the second derivative curves. For instance, this could result in a less pronounced structure for the first phonon branch (acoustic modes) in $\alpha^2F(\omega)$ compared to the second derivative data of the junctions. All this also influences the resulting parameters for $\lambda$, $\mu^*$, and $T_c$. Moreover, the presence of the (minor) spurious effects cannot be excluded at low magnetic fields.

Our results for the electron-phonon interaction (Fig.\ref{Fig5}) can be compared with the inversion of the tunneling data on $\rm Ba_{1-x}K_xBiO_3$ by Huang \emph{et al}. \cite{7} and the theoretical calculation of the Eliashberg function of $\rm Ba_{1-x}K_xBiO_3$ by Shirai, Suzuki, and Motizuki \cite{12}. In the data of Huang \emph{et al}. two dominant peaks are clearly observed up to 50~$meV$ with the minimum in between at 20~$meV$. In our spectra, the first branch of the spectrum (up to 20~$meV$) is not resolved and overlaps with the second one. Because structure is observed in this region in our measured second derivative $d^2V /dI^2$, the absence of a clear maximum for the first branch could be related to the above mentioned uncertainties in the reduced density of states in this frequency range. A detailed comparison with the data of Huang et \emph{al}.\cite{7} shows a remarkable agreement in magnitude and shape of $\alpha^2F$ in the energy range 25-55. Our data, in contrast to those of Huang \emph{et al}., also reveal a significant spectral contribution in the high-energy range around 60~$meV$. The most pronounced feature which can be found in our Eliashberg function is a minimum in the spectrum at 50~$meV$. This minimum can be found in the phonon density of states $F(\omega)$ as well as in the Eliashberg function calculated by Shirai, Suzuki, and Motizuki  \cite{12}. Our spectra therefore demonstrate the importance of the high-frequency optical modes of oxygen in the electron-phonon interaction, in agreement with first-principles calculations \cite{12}.

\section{CONCLUSIONS}
Low leakage tunnel junctions on $\rm Ba_{1-x}K_xBiO_3$ have been investigated both in the superconducting and in the normal state. The effect of the magnetic field on the gap-like peaks in the tunneling conductance has been measured up to 10~$T$.

The normalized tunneling conductance was successfully fitted to the smeared BCS density of states with the resulting superconducting energy gap $\Delta=5\ meV$, the lifetime broadening $\Gamma=1.7\ meV$ and the ratio $2\Delta /k_BT_c$ equal to 4.1. Because in tunneling measurements of other workers \cite{7,8,9} the ideal BCS gap structure has been observed without any smearing, we ascribe the smearing effect to an extrinsic origin such as inhomogeneous stoichiometry.

A reproducible structure was observed at characteristic phonon energies in the measured second derivative $d^2V/dI^2(V)$. This structure is often interfering with spurious (current-driven) effects. The spurious signal can be suppressed by the application of a magnetic field leaving a clear phonon structure in the $d^2V/dI^2$ spectra. The McMillan-Rowell inversion of the tunneling data yields the Eliashberg interaction function $\alpha^2F(\omega)$ which resembles the phonon density of states obtained by neutron scattering experiments \cite{13} as also the Eliashberg function calculated theoretically by Shirai, Suzuki, and Motizuki \cite{12}. The spectral weight in the high-frequency region points to an important contribution of the optical modes of oxygen in the electron-phonon interaction in agreement with the first-principle calculation \cite{12}. The obtained parameters $\lambda$ and $\mu^*$ give the transition temperature $T_c$ close to the experimental $T_c =28\ K$.

It is desirable to perform similar measurements on more perfect samples with no broadening in the BCS quasiparticle spectrum. It would be also interesting to perform the tunneling experiment on a thin film (film thickness smaller than the London penetration depth) in a parallel magnetic field with no vortices inside the sample. Similar experiments have been performed by Rasing \emph{et al}. \cite{19} on thin lead films. Nevertheless, our measurements support the important role of the electron-phonon interaction in the superconductivity of $\rm Ba_{1-x}K_xBiO_3$.
\section{ACKNOWLEDGMENTS}
We are very grateful to Professor S.I. Vedeneev and Dr. S.V. Meshkov for their help in handling the observed broadening in the superconducting density of states in the analysis of the data.

\end{document}